\documentclass{ws-ijgmmp}
\usepackage{color}

\DeclareMathOperator{\grad}{grad}

\DeclareMathOperator*{\Ric}{Ric}

\newcommand{\D}{\partial}

\begin{document}

\markboth{F. Gholami, A. Haji-Badali, F. Darabi}
{CLASSIFICATION OF EINSTEIN EQUATIONS WITH COSMOLOGICAL CONSTANT IN WARPED PRODUCT SPACE-TIME}

%
\catchline{}{}{}{}{}
%

\title{CLASSIFICATION OF EINSTEIN EQUATIONS WITH COSMOLOGICAL CONSTANT IN WARPED PRODUCT SPACE-TIME}

\author{F. Gholami}

\address{Department of Mathematics, Basic Sciences Faculty,\\
   University of Bonab, Bonab, Iran.\\
              \email{fateme.gholami@bonabu.ac.ir} }

\author{A. Haji-Badali}

\address{Department of Mathematics, Basic Sciences Faculty,\\
   University of Bonab, Bonab, Iran.\\
              \email{haji.badali@bonabu.ac.ir} }

\author{F. Darabi\footnote{
Corresponding author.}}

\address{Department of Physics,
Azarbaijan Shahid Madani University, Tabriz, Iran.
\email{f.darabi@azaruniv.edu} }

\maketitle

\begin{history}
\received{(Day Month Year)}
\revised{(Day Month Year)}
\end{history}

\begin{abstract}
We classify all warped product space-times in three categories as i) generalized twisted product structures, ii) base conformal warped product structures and iii) generalized static space-times and then we obtain the Einstein equations with the corresponding cosmological constant by which we can determine uniquely the warp functions in these warped product space-times. 
\end{abstract}


\section{Introduction}

Twisted product space-times are generalizations of warped product space-times
in that the warping function may depend on the points of both warp factors \cite{Fernandez,PR}.  Base conformal warped product space-times have metrics in the form of a mixture of a conformal metric on the base and a warped metric. These metrics with considerations about their curvatures are very frequent in different aspects of physics, such as relativity theory, extra-dimensional theories (Kaluza-Klein, Randall-Sundrum), string and super-gravity theories, quantum gravity and also the study of spectrum of Laplace-Beltrami operators on $p$-forms. On the other hand, a standard static space-time is a Lorentzian warped product space-time where the warp function is defined on a Riemannian manifold and acts on the negative definite metric on an open interval of the real numbers. Different aspects of these space-times have been previously studied on many issues, such as geodesic equation, geodesic completeness, geodesic connectedness, causal structure, curvature conditions, etc (see for example \cite{Anderson.Chrusciel.Delay,bejancu.farran,bejancu.farran:book,Dobarro.unal2}).

On the other hand, the Einstein equations are considered as the basic equations
in General Relativity. In fact, finding the Einstein equations is of prime
importance in any general relativistic model of gravitation and cosmology.
In recent years, a large amount of attention has been paid on the
modification of Einstein equations \cite{f(R)}, specially in models with space-time dimensions other than four \cite{RSDGP}. In this line of activity, we have previously obtained the Einstein's equation in $(m+n) D$ and $(1+n) D$ multidimensional space-time with multiply-warped product metric $(\bar{M},\bar{g})$ \cite{MF-AH-FGH}, \cite{FGH-FD-AH}. Specially, we have discussed on the origin of $4D$ cosmological constant as an emergent effect of higher dimensional warped spaces.

In this work, we intend to obtain the Einstein's equations with cosmological constant in the warped product space-times categorized as i) generalized twisted product structures, ii) base conformal warped product structures and iii)
generalized static space-times. In the beginning, we recall that a pseudo-Riemannian manifold $(M_1, g)$ is conformal to the pseudo-Riemannian manifold $(M_2, h)$, if and only if there exists $\eta \in C^\infty(M_1)$ such that $g = e^\eta
h$. Also, we will call a doubly twisted product as a base conformal warped
product, when the functions $\psi$ and $\varphi$ depend only on the
points of $ M_1$.

\section{Preliminaries}

 \begin{defi}
Let $(M_1,g)$ and $(M_2,h)$ are two pseudo-Riemannian manifolds,
where $M_1$ is a $m$-dimensional manifold and  $M_2$ is a
$n$-dimensional manifold and $f$ be a positive smooth function on $
M_1$. The local coordinates on $\bar{M}=M_1\times{}_fM_2$ are $(
x^i,x^\alpha)$, where $(x^i)$  and $(x^\alpha)$ are the local
coordinates on $ M_1$ and $ M_2$, respectively. Also suppose that,
$\bar{g}$ is a pseudo-Riemannian metric on $\bar{M}$ which is
defined by
\begin{align}
 \bar{g} = g_{ij}dx^idx^j + f^2(x^k)h_{\alpha\beta}dx^\alpha
 dx^\beta,\label{eq: metric1}
 \end{align}
 which is given by its local components as:
 \begin{align}
 \bar{g}_{ij}=g_{ij},\ \   \bar{g}_{\alpha\beta}= f^2(x^k)h_{\alpha\beta},\ \
 \bar{g}_{i\alpha}=0,\label{eq:metric2}
 \end{align}
 where $g_{ij}(x^k)$ are the local components of $g$ and $h_{\alpha\beta}(x^\mu)$ are the local components of $h$.
 Here, and in the sequel, we use the following ranges for indices:
$ i, j, k, ...\in \{1,...,m\}$; $\alpha, \beta, ...\in \{m+1,...,m+n\}$;
  $a,b,c,... \in \{1,...,n+m\}$.

  \end{defi}

We denote by $\bar{G}$ the Einstein gravitational tensor field of
$(\bar{M}, \bar{g})$, that is, we have,
\begin{align}
\bar{G} = \bar{\Ric} - \frac{1}{2}\bar{S}\bar{g},\label{eq:Einstein}
\end{align}
where $\bar{\Ric}$ and $\bar{S}$ are Ricci tensor and scaler curvature of $\bar{M}$, respectively.

\section{Generalized twisted product structures}
Twisted products are generalizations of warped products, namely the
warping function may depend on the points of both factors.  Twisted
product are useful in the study of Einstein's equations, when we use
 special conditions of this products.

  Let $(M_1,g)$ and $(M_2,h)$ be two pseudo-Riemannian manifolds
and let $f : M_1\times M_2\rightarrow\mathbb{R^+}$  be a positive
function on $M_1\times M_2$. The product manifold $\bar{M} =
M_1\times M_2$ endowed with the metric $\bar{g} = g\oplus f^2h$ is
called a twisted product. The same terminology of base and fiber
applies in this case, whereas $f$ is the twisting function
(sometimes
it is also referred to as the warping function).\\
For simplicity in some of the expressions, we use $\xi = Log f$
instead of $f$\\
\begin{pro}[\cite{Fernandez}]
 Let $\bar{M} =
M_1\times_f M_2$ be a twisted product we have
\begin{align}
\bar{Ric}(\partial i,\partial j )& =Ric^{M_1}(\partial i,\partial j) - n (\partial i(\xi)\partial j(\xi) + H_\xi(\partial i,\partial j)),\nonumber\\
\bar{Ric}(\partial i,\partial \alpha ) & = (1 - n)\partial i
\partial\alpha  (\xi),\\
\bar{Ric}(\partial \alpha,\partial \beta )& = Ric^{M_2}(\partial
\alpha,\partial \beta)  + (2 - n)(\partial \alpha(\xi)\partial
\beta(\xi)+ H_\xi(\partial \alpha,\partial \beta))\nonumber\\ & +
(n-2)\bar{g}(\partial \alpha,\partial
\beta)\bar{g}(\nabla\xi,\nabla\xi) - \bar{g}(\partial
\alpha,\partial \beta)\Delta\xi,\nonumber
\end{align}
 where $H$ is the Hessian tensor of  $\xi$.
\end{pro}

 Now let $(\bar{M},\bar{g})$ be a    space-time,
where $\bar{M}=M\times_f I$ is  a $(n+1)$- dimensional twisted
product manifold, $M$ is a $n$-dimensional manifold, $g$ is a
Lorentz metric on $M$, $I$ is a 1-dimensional manifold and $f :
M\times I\rightarrow\mathbb{R^+}$  be a positive function on
$M\times I$. Also, suppose that $ \bar{ g}$ is a semi-Riemannian
metric on $\bar{M}$, given by its local components
\begin{align}
\bar{g} (\partial i,\partial j)& = g(\partial
i,\partial j),\nonumber \\
\bar{g}(\partial i,\frac{\partial t}{f})& = 0, \\
\bar{g}(\frac{\partial t}{f},\frac{\partial t}{f})& = f^2,\nonumber
\end{align}
where $g(\partial \alpha,\partial \beta)$ are the local components
of $g$. Hence, the metric defined by  $\bar{g}$ has the form
\begin{align}
 ds^2=g(\partial i,\partial j) dx^i dx^j +
f^2(dt)^2.
\end{align}
\begin{theorem}[\cite{Fernandez}]
 Let $\bar{M} = M_1\times_f M_2$ be a twisted product of $(M_1, g)$ and $(M_2, h )$ with
twisting function $f$ and ${\rm dim}  M_2 > 1$. Then, $Ric(\partial
i,\partial \alpha ) = 0$ if and only if $M_1\times_f M_2$ can be
expressed as a warped product, $M_1\times_\psi M_2$ of  $(M_1, g)$
and $(M_2, \tilde{h} )$ with a warping function $\psi$, where
$\tilde{h}$ is a conformal metric tensor to $h$.
\end{theorem}
\begin{pro}[\cite{Chen}]
Let $\bar{M}=M\times_f I$ be a twisted product. Let $\partial
i,\partial j $  are local components of vectors tangent to the base
$M$ and let $\frac{\partial t}{f}$ be  local components of vector
tangent to the fiber. 
 Then the Ricci tensor is:
\begin{align}\label{7}
\bar{Ric}(\partial i,\partial j )&=Ric^M(\partial i,\partial j )
-\frac{H_f (\partial i,\partial
j)}{f}, \nonumber\\
\bar{Ric}(\partial j,\frac{\partial t}{f} )&=0,\\
\bar{Ric}(\frac{\partial t}{f},\frac{\partial t}{f}
)&=-\frac{\Delta^M f}{f},\nonumber
\end{align}
where $Ric^{M}(\partial_i,\partial_j)$ are the local components of
Ricci tensor  on $(M,g)$.
\end{pro}

Now, in what followed, according a straightforward calculation, we obtain Einstein equations  and cosmological constant on
$\bar{M}=M\times_f I.$\\
\begin{pro}\label{pro1}
The Einstein gravitational tensor field of $(\bar{M},\bar{g})$, have
following equations:
\begin{align}\label{8}
\bar{G}(\partial_i,\partial_j)&=G(\partial_i,\partial_j)-\frac{H_f
(\partial i,\partial j)}{f}+\frac{\Delta^M f}{f}g(\partial
i,\partial j),\nonumber\\
\bar{G}(\frac{\partial t}{f},\frac{\partial t}{f})&=-\frac{1}{2}f^2
S^M - \frac{\Delta^M f}{f} + f\Delta^M
f ,\nonumber\\
\bar{G}(\frac{\partial t}{f},\partial_i)&=0,\nonumber\\
\end{align}
\end{pro}
\begin{proof}
By the equation (\ref{eq:Einstein}), and a straightforward computation according the argument in (\ref{7}) we will have the desired result.
\end{proof}

Let $(\bar{M},\bar{g})$ be the twisted product space. Suppose that the Einstein gravitational tensor field
$\bar{G}$ of $(\bar{M},\bar{g})$ satisfies the
Einstein equations with cosmological  constant $\bar{\Lambda}$ as
\begin{align}
\bar{G}=-\bar{\Lambda}\bar{g}.\label{6.1}
\end{align}
Now according to the Proposition \ref{pro1} and a straightforward computation to Einstein equation $\bar{G} =
-\bar{\Lambda}\bar{g}$, we have the following theorem.
\begin{theorem}
The Einstein equations on $(\bar{M},\bar{g})$ with cosmological
$\bar{\Lambda}$ are equivalent with the following equations
\begin{align}
\bar{\Lambda}&=\frac{\Delta^M
f}{f}(\frac{1}{f^2}(1-\frac{n}{2})-\frac{1}{2}),\nonumber\\
G(\partial_i,\partial_j)&=\frac{H_f(\partial i,\partial
j)}{f}-\frac{\Delta^M
f}{f}(\frac{1}{f^2}(1-\frac{n}{2})-\frac{1}{2})g(\partial i,\partial
j).
\end{align}
\end{theorem}
 Now, by demanding for a constant $\bar{\Lambda}$, the warp function $f$ is obtained for a given dimension $n$, as a
 solution of the following differential equation

\begin{equation}
\frac{\Delta^M f}{f}(\frac{1}{f^2}(1-\frac{n}{2})-\frac{1}{2})=a,
\end{equation}

where $a$ is a constant which can take negative, zero and positive values
corresponding to negative, zero and positive cosmological constants.

\section{ Special base conformal warped product }
Doubly-twisted product is a usual product of  pseudo-Riemannian
manifolds $ M_1\times M_2$ with $g = f_1^2g \oplus f_2^2 h$, then
$\bar{M} = M_1\times_{(f_1,f_2)} M_2$ is called as
\textit{doubly-twisted product}. When $f_1=1 $ then $ M_1\times_{f}
M_2$ is a twisted product which we discussed in  the last section.
Doubly twisted product is  as a base conformal warped product when
the functions $f_1$  and $f_2$ only depend on the points of $M_1$.
In this section we  consider a subclass of base conformal warped
product called as special base conformal warped products.
 \begin{defi}
Let  $(M_1,g )$   and $(M_2,h )$ be $m$ and $n$ dimensional pseudo
Riemannian manifolds, respectively. Then $\bar{M} = M_1\times M_2$
is an $(m + n)$-dimensional pseudo Riemannian manifold with smooth
functions $f_1:M_1\rightarrow (0,\infty) $  and
$f_2:M_1\rightarrow(0,\infty)$. The base conformal warped product
is the product manifold $\bar{M} = M_1\times M_2$ furnished with the
metric tensor $\bar{g} = f_1^2g \oplus f_2^2 h$ defined by
\begin{align}
 \bar{g} = (f_1\circ \pi)^2\pi^*g \oplus (f_2 \circ \sigma)^2 \sigma^*h.
\end{align}
We will denote this structure by $M_1\times_{(f_1,f_2)} M_2$. The
function $f_2:M_1\rightarrow (0,\infty) $ is called the warping
function and the function $f_1:M_1\rightarrow(0,\infty)$ is said to
be the conformal factor.

If $f_1 = 1$ and $f_2$ is not identically 1, then we obtain a singly
warped product. If both $f_1 = 1$ and $f_2 =1$, then we have a
product manifold. If neither $f_1$ nor $f_2$ is constant, then we
have a nontrivial base conformal warped product. If $(M_1,g )$ and
$(M_2,h )$ are both Riemannian manifolds, then
$M_1\times_{(f_1,f_2)} M_2$ is also a Riemannian manifold. We call
$M_1\times_{(f_1,f_2)} M_2$ as a Lorentzian base conformal warped
product if $(M_2,h )$ is Riemannian and either $(M_1,g )$ is
Lorentzian or else $(M_1,g )$ is a one-dimensional manifold with a
negative definite metric $-dt_2$.
\end{defi}

\begin{pro}[\cite{Dobarro.unal1}]
Let $\partial i, \partial j $ and $\partial \alpha,\partial \beta$
are local components of vectors tangent to the base $M_1$ and fiber
$M_2$, respectively. Then, the Ricci
tensor of $M_1\times_{(f_1,f_2)} M_2$ satisfies\\
\begin{align}
Ric(\partial i,\partial j)&= Ric^{M_1}(\partial i,\partial j) - (m -
2)\frac{1}{f_1} H_{M_1}^{f_1}(\partial i,\partial j)
+ 2(m - 2)\frac{1}{f_1^2}\partial i(f_1)\partial j(f_2)\nonumber\\
&-[(m-3)\frac{g(\nabla^{M_1}f_1,\nabla^{M_1}f_1)}{f^2_1}+\frac{\Delta_{M_1}{f_1}}{f_1}]g(\partial
i,\partial j)-n\frac{1}{f_2}
H_{M_1}^{f_2}(\partial i,\partial j)\nonumber\\
&-n\frac{g(\nabla^{M_1}{f_2},\nabla^{M_1}{f_1})}{f_1f_2}g(\partial
i,\partial j)+n
\frac{\partial i(f_1)}{f_1}\frac{\partial j(f_2)}{f_2} + n\frac{\partial j(f_1)}{f_1}\frac{\partial i(f_2)}{f_2},\nonumber \\
Ric(\partial i,\partial\alpha)&= 0,\\
Ric(\partial \alpha,\partial \beta)& = Ric^{M_2}(\partial
\alpha,\partial \beta) -\frac{f_2^2}{f_1^2}h(\partial
\alpha,\partial \beta)[(m -
2)\frac{g(\nabla^{M_1}f_2,\nabla^{M_1}{f_1})}{f_1f_2} +
\frac{\Delta_{M_1}{f_2}}{f_2}\nonumber \\
&+(n
-1)\frac{g(\nabla^{M_1}{f_2},\nabla^{M_1}{f_2})}{f_2^2}].\nonumber
\end{align}
\end{pro}

\begin{pro}[\cite{Dobarro.unal1}]
The scalar curvature S of $M_1\times_{(f_1,f_2)} M_2$ is given by\\
\begin{align}
f_1^2S&=S_{M_1} + S_{M_2}\frac{f_1^2}{f_2^2} - 2(m - 1)\frac{
\Delta_{M_1}{f_1}}{f_1} - 2n\frac {\Delta_{M_1}{f_2}}{f_2}-(m -
4)(m -1)\frac{g(\nabla^{M_1}{f_1},\nabla^{M_1}{f_1})}{f_1^2}\nonumber\\
&-2n(m - 2)\frac{g(\nabla^{M_1}{f_2},\nabla^{M_1}{f_1})}{f_1f_2}-n(n
- 1) \frac{g(\nabla^{M_1}{f_2},\nabla^{M_1}{f_2})}{f_2^2}.
\end{align}
\end{pro}

\subsection{Generalized Robertson-Walker (GRW) space-time}

\begin{defi}
{\color{red}A} $(n+1)$-dimensional generalized Robertson-Walker (GRW) space-time
with $n>1$ is a Lorentzian manifold which is  the base conformal
warped product $\bar{M}=I\times_{(f_1,f_2)} M$ of an open interval
$I$ of the real line $\mathbb{R}$ and a Riemannian $n$-manifold $(M,
g)$ endowed with the Lorentzian metric

\begin{align}
\bar{g}=-f_1(t)^2\pi^*(dt^2)+f_2(t)^2\sigma^*(g),
\end{align}

we can write
\begin{align}
\bar{g}=-dt^2+f_2(t)^2h_{\alpha\beta}dx^\alpha dx^\beta,
\end{align}
where $\pi$ and $\sigma$ denote the projections onto $I$ and $M$,
respectively, and $f_1$, $f_2$ are  positive smooth function on $I$.
In a classical Robertson–Walker  space-time, the fiber is three
dimensional and of constant sectional curvature, and the warping
function $f_1$ and  $f_2$ are arbitrary. The following formula can
be directly obtained from the previous result and noting that on a
multiply generalized Robertson–Walker space-time $\grad_I f =
-f',\Vert \grad_If\Vert^2_I =-f'^2, g(\frac{\D}{\D t}, \frac{\D}{\D
t})=-1,H^f(\frac{\D}{\D t}, \frac{\D}{\D t})=f''$
 and $\Delta_I f=-f''$, we denote the usual derivative
on the real interval $I$ by the prime notation (i.e.,$'$)  from now
on.
 \end{defi}
 \begin{pro}[\cite{oneil:book}]
Let $\bar{M}=I\times_{(f_1,f_2)} M_2$ be {\color{red}the} base conformal warped
product manifold.
\begin{align}
\bar{\Ric}(\partial_t,\partial_t)&=-n\frac{f_2''}{f_2},
\nonumber\\
\bar{\Ric}(\partial_t,\partial_\alpha)&=0,\label{eq:Ric2}\\
\bar{\Ric}(\partial_\alpha,\partial_\beta)&=Ric^{M_2}(\partial_\alpha,\partial_\beta)-f_2^2g(\partial_\alpha,\partial_\beta)
\left(-\frac{f_2''}{f_2}{\color{red}-}(n-1)(\frac{f_2'}{f_2})^2\right)
,\nonumber
\end{align}
where ${\color{red}Ric^{M_2}}(\partial_\alpha,\partial_\beta)$ are
the local components of  Ricci tensor in $(M_2,h)$.
\end{pro}
\begin{pro}[\cite{oneil:book}]
Let $\bar{M}=I\times_{(f_1,f_2)} M_2$ be  the base conformal warped
product manifold. Then scalar curvature $S$ of $(\bar{M},\bar{g})$
admits the following expression,
\begin{align}
\bar{S}=\frac{S_{M_2}}{f_2^2}+2n\frac{f_2''}{f_2}{\color{red}+}n(n-1)(\frac{f_2'}{f_2})^2.
\end{align}
\end{pro}
So we will have:
\begin{pro}
Let $\bar{G}$ be the Einstein gravitational tensor field of
$(\bar{M},\bar{g})$, then we have following equations:\\
\begin{align}
\bar{G}(\partial_t,\partial_t)&=-\frac{1}{2}(-\frac{S_{M_2}}{f_2^2}-n(n-1)(\frac{f_2'}{f_2})^2),\nonumber\\
\bar{G}(\partial_t,\partial_\alpha)&=0,\\
\bar{G}(\partial_\alpha,\partial_\beta)&=G(\partial_\alpha,\partial_\beta)-f_2^2g(\partial_\alpha,\partial_\beta)
\left((n-1)\frac{f_2''}{f_2}-(n-1)(1-\frac{n}{2})(\frac{f_2'}{f_2})^2\right),\nonumber
\end{align}
where $G(\partial_\alpha,\partial_\beta)$ are the local components
of the Einstein gravitational tensor field $G$ of $(M_2, h)$ and
$G(\partial_t,\partial_t)$ are the local components of the Einstein
gravitational tensor field $G$ of $(I, g)$.
\end{pro}

%

\begin{theorem}
The Einstein equations on $(\bar{M},\bar{g})$ with cosmological
$\bar{\Lambda}$ are equivalent with the following equations
\begin{align}
\bar{\Lambda}&=\frac{1}{2}n(n-1)\frac{f_2''}{f_2},\\
G(\partial_\alpha,\partial_\beta)&=f_2^2g(\partial_\alpha,\partial_\beta)(n-1)(\frac{n}{2}-1)\left(\frac{f_2''}{f_2}-(\frac{f_2'}{f_2})^2\right).\nonumber
\end{align}

\end{theorem}
 Now, by demanding for a constant $\bar{\Lambda}$, the warp function $f$ is obtained for a given
  dimension $n$, as the unique solution of the following differential equation
\begin{equation}
\frac{1}{2}n(n-1)\frac{f_2''}{f_2}=a,
\end{equation}
where $a$ is a constant which can take negative, zero and positive values
corresponding to negative, zero and positive cosmological constants.

%

\section{Generalized standard static space-times}
 In this section we study a standard static space-time 
which is
$n$-dimensional  Lorentzian product. It is a generalization of a
Einstein static universe. The results  are important in solutions of
the
Einstein equations. Partial results are physical motivation for cosmological constant problems.

 Let $(M,g)$ be an $n$-dimensional
Riemannian manifold and $f: M \rightarrow (0,\infty)$ be a smooth
function. Then $(n+1)$-dimensional product manifold $\bar{M}=(a,
b)\times M $ furnished with the metric tensor $\bar{g} =-f^2dt^2
\oplus g_M$ is called a standard static space-time and is denoted by
$_f(a, b)\times M$, where $dt^2$ is the Euclidean metric tensor on
$(a,
b)$ and $-\infty\leq a \leq b\leq\infty$.

  We have in local components
\begin{align}
\bar{g} (\partial \alpha,\partial \beta)& = g(\partial \alpha,\partial \beta),\nonumber \\
\bar{g}(\partial \alpha,\frac{\partial t}{f})& = 0, \\
\bar{g}(\frac{\partial t}{f},\frac{\partial t}{f})& = -f^2.\nonumber
\end{align}
\begin{pro} [\cite{Dobarro.unal2}]
Let  $_f(a, b)\times M$ be a standard static space-time. We have
following equation for scalar curvatures of the space-time
\begin{align}
\bar{S}=S_M - 2\frac{\Delta^M(f)}{f},
\end{align}
 where Ricci tensor have following equations

\begin{align}
\bar{\Ric}(\frac{\partial t}{f},\frac{\partial
t}{f})&=\frac{\Delta^M(f)}{f},
\nonumber\label{eq:ric1}\\
\bar{\Ric}(\frac{\partial t}{f},\partial_\alpha)&=0,\\
\bar{\Ric}(\partial_\alpha,\partial_\beta)&=Ric^M(\partial_\alpha,\partial_\beta)-\frac{H_M^f(\partial_\alpha,\partial_\beta)}{f}
.\nonumber
\end{align}
\end{pro}
So we easily deduce the following:
\begin{pro}
the Einstein gravitational tensor field of
$(\bar{M},\bar{g})$, have following equations:
\begin{align}
\bar{G}(\frac{\partial t}{f},\frac{\partial t}{f})&=\frac{1}{2}f^2
S^M + \frac{\Delta^M f}{f} - f\Delta^M
f ,\nonumber\\
\bar{G}(\partial\alpha,\partial\beta)&=G(\partial\alpha,\partial\beta)-\frac{H_f
(\partial \alpha,\partial \beta)}{f}+\frac{\Delta^M f}{f}g(\partial
\alpha,\partial \beta),\nonumber\\
\bar{G}(\frac{\partial t}{f},\partial_i)&=0,\nonumber\\
\end{align}
\end{pro}
%
\begin{theorem}
The Einstein's equations on $(\bar{M},\bar{g})$ with cosmological
$\bar{\Lambda}$ are expressed with the following equations

\begin{align}
\bar{\Lambda}&=\frac{\Delta^M
f}{f}(\frac{1}{f^2}(1-\frac{n}{2})-\frac{1}{2}),\nonumber\\
G(\partial\alpha,\partial\beta)&=\frac{H_f(\partial \alpha,\partial
\beta)}{f}-\frac{\Delta^M
f}{f}(\frac{1}{f^2}(1-\frac{n}{2})-\frac{1}{2})g(\partial
\alpha,\partial \beta).
\end{align}
\end{theorem}

 Now, by demanding for a constant $\bar{\Lambda}$, the warp function $f$ is obtained for a given dimension $n$, as
 a solution of the following differential equation
\begin{equation}
\frac{\Delta^M f}{f}(\frac{1}{f^2}(1-\frac{n}{2})- \frac{1}{2})=a,
\end{equation}
where $a$ is a constant which can take negative, zero and positive values
corresponding to negative, zero and positive cosmological constants.

%

\section{Conclusions}
In this paper, we have studied three classes of warped product space-times as: generalized twisted product structures, base conformal warped product structures and generalized static space-times. Then, we obtained Einstein equations with cosmological constant in these warped product space-times. We may conclude that any Einstein equations with cosmological constant in warped product space-times lies within these three categories. Therefore, a variety of gravitational and cosmological models with warped product structure can be categorized according to the corresponding warp functions.


\begin{thebibliography}{0}


\bibitem{Fernandez}
 M. Fernández-López, E. Garc\'{i}a-R\'{i}o, D. N. Kupeli and B. \"{U}nal, A curvature condition for a twisted
product to be a warped product, {\it manuscripta mathematica}, {\bf106}(2)
(2001), 213--217.
\bibitem{PR}
 R. Ponge  and R. Helmut,  Twisted products in pseudo-Riemannian geometry, {\it Geometriae Dedicata}, {\bf 48}(1) (1993), 15--25.
\bibitem{Anderson.Chrusciel.Delay}
M. T. Anderson, T. P. Chrusciel and E. Delay,
 Non-trivial, static, geodesically complete, vacuum space-times with
a negative cosmological constant, {\it Journal of High Energy Physics},
{\bf 2002} (10), (2002): 063.

\bibitem{bejancu.farran}
A. Bejancu, C. Constantin, and H.~R.  Farran,
Classification of 5d warped spaces with cosmological constant, {\it Journal of Mathematical Physics}, {\bf 53}, (2012), 122503.

\bibitem{bejancu.farran:book}
A. Bejancu and H.~R. Farran,   Foliations and geometric structures, vol.~580 of {\it
Mathematics and Its Applications (Springer)}.
 Springer, Dordrecht, 2006.
\bibitem{Dobarro.unal2}F. Dobarro and B.  \"{U}nal. Special standard static
space--times, {\it Nonlinear Analysis: Theory, Methods and Applications},
{\bf 59}(5) (2004), 759--770.
\bibitem{f(R)}S. Capozziello and M. Francaviglia, Extended Theories of Gravity and their Cosmological and Astrophysical Applications, Gen. Rel. Grav.40, (2008), 357.
\bibitem{f(R)1}T. P. Sotiriou and V. Faraoni, $f(R)$ Theories Of Gravity,
 Rev. Mod. Phys.82, (2010), 451.
\bibitem{f(R)2}S. Capozziello and M. De Laurentis, {\it Invariance Principles and Extended Gravity: Theory and Probes}, (Nova Science Publishers, 2010).
\bibitem{f(R)3}S. Capozziello, Curvature Quintessence, Int. J. Mod. Phys. D11, (2002) 483.
\bibitem{f(R)4}D. Lovelock, The Einstein Tensor and Its Generalizations, J. Math. Phys., 12 (3), (1971), 498.
\bibitem{f(R)5}S. Nojiri and S. D. Odintsov, Where new gravitational physics comes from: M-theory?, Phys. Lett. B576, (2003), 5.
\bibitem{f(R)6}S. Nojiri and S. D. Odintsov, Modified gravity with negative and positive powers of the curvature: unification of the inflation and of the cosmic acceleration, Phys. Rev. D68, (2003), 123512.
\bibitem{f(R)7}S. M. Carroll, V. Duvvuri, M. Trodden, and M. S. Turner, Dark Energy and the Accelerating Universe, Phys. Rev. D70, (2004), 043528.
\bibitem{f(R)8}G. Allemandi, A. Borowiec, and M. Francaviglia, Accelerated Cosmological Models in Ricci squared Gravity, Phys. Rev. D70, (2004), 103503.
\bibitem{f(R)9}S. Capozziello, V. F. Cardone, and A. Troisi, Reconciling dark energy models with f(R) theories, Phys. Rev. D71, (2005), 043503.
\bibitem{f(R)10}S. Carloni, P. K. S. Dunsby, S. Capozziello, and A. Troisi, Cosmological dynamics of $R^n$ gravity, Class. Quant. Grav.22, (2005), 4839.
\bibitem{f(R)11}P. K. S. Dunsby, E. Elizalde, R. Goswami, S. Odintsov, and D. Saez-Gomez, On the LCDM Universe in $f(R)$ gravity, Phys. Rev. D82, (2010), 023579.
\bibitem{f(R)12}N. Goheer, J. Larena, P. K. S. Dunsby, Power-law cosmic expansion in f (R) gravity models, Phys. Rev. D80, (2009), 061301.
\bibitem{f(R)13}S. Nojiri and S. D. Odintsov, Unified cosmic history in modified gravity: From $F(R)$ theory to Lorentz non-invariant models Phys. Rept.505,  (2011), 59.
\bibitem{f(R)14}S. Nojiri and S. D. Odintsov, introduction to modified gravity and gravitational alternative for dark energy, Int. J. Geom. Meth. Mod. Phys.4, (2007), 115.
\bibitem{f(R)16}A. De Felice and S. Tsujikawa, $f(R)$ Theories, Living Rev. Rel.13, (2010), 3.
\bibitem{f(R)17}Yi-Fu Cai, S. Capozziello, M. De Laurentis, E. N. Saridakis, $f(T)$ teleparallel gravity and cosmology, Rept. Prog. Phys. 79 (2016), 106901.
\bibitem{RSDGP}L. Randall, and R. Sundrum, Large Mass Hierarchy from a Small Extra Dimension, Physical Review Letters 83.17 (1999), 3370.
\bibitem{RSDGP1}L. Randall, and R. Sundrum, An Alternative to Compactification,
Physical Review Letters 83.23 (1999), 4690.
\bibitem{RSDGP2}G. Dvali, G. Gabadadze, and M. Porrati, 4D Gravity on a Brane in 5D Minkowski Space, Physics Letters. B485 (2000), 208.
\bibitem{MF-AH-FGH}M. Faghfouri,  A. Haji-Badali,  and F. Gholami, On cosmological constant of Generalized Robertson-Walker space-times, J. Math. Phys 58, (2017) 053508.

\bibitem{FGH-FD-AH}F. Gholami, F. Darabi and A. Haji-Badali, Multiply-warped product metrics and reduction of Einstein equations,{ Int. J. Geometric. M. Modern. Physics}. {\bf 14}(2) (2017), 1750021 (11 pages).
\bibitem{Chen}B.-Y. Chen, Geometry of submanifolds and its applications,
{\it Science University of Tokyo}, 1981.

\bibitem{Dobarro.unal1}F. Dobarro  and B. \"{U}nal, Curvature in special base conformal warped products, {\it Acta Applicandae Mathematicae}, {\bf 104} (1) (2008), 1-46.

\bibitem{oneil:book}
B. O'Neill,  Semi-{R}iemannian geometry with applications to relativity,
vol.~103 of {\em Pure and Applied Mathematics}.  Academic Press Inc., New York, 1983.

\end{thebibliography}
\end{document}